\documentstyle[sprocl]{article}

\def\NPB#1#2#3{{\it Nucl.\ Phys.}\/ {\bf B#1} (#2) #3}
\def\PLB#1#2#3{{\it Phys.\ Lett.}\/ {\bf B#1} (#2) #3}
\def\PRD#1#2#3{{\it Phys.\ Rev.}\/ {\bf D#1} (#2) #3}

\def\be{\begin{equation}}
\def\ee{\end{equation}}
\def\bea{\begin{eqnarray}}
\def\eea{\end{eqnarray}}

\def\Tr{{\rm Tr}}

\begin{document}

\title{THREE-FAMILY PERTURBATIVE STRING VACUA: FLAT DIRECTIONS AND EFFECTIVE
COUPLINGS~\footnote{Talk given at PASCOS 98, Northeastern University, Boston, MA,
March 1998.}
}

\author{LISA EVERETT}

\address{Department of Physics and Astronomy\\ University of Pennsylvania,
Philadelphia, PA 19104, USA\\E-mail: lle@sas.upenn.edu}

\maketitle\abstracts{The properties of a class of quasi-realistic three-family
perturbative heterotic string vacua are addressed.  String models in this class 
generically contain an anomalous $U(1)$, such that the nonzero Fayet-Iliopoulos
term triggers certain fields to acquire string scale VEV's along flat directions.    
This vacuum shift reduces the rank of the gauge group and generates
effective mass terms and effective trilinear interactions.  Techniques are
discussed which yield a systematic classification of the flat directions of a 
given string model which can be proven to be $F$- flat to all orders. The
effective superpotential along such flat directions can then be calculated to
all orders in the string (genus) expansion.}

\section{Introduction}
There are several challenges to be faced in the investigation of the phenomenology
of string models, including the degeneracy of string models (but as yet no fully
realistic one) and the absence of a satisfactory scenario for supersymmetry
breaking.  However, a class of quasi-realistic models have been constructed in
perturbative heterotic string theory, particularly in the free fermionic
constructions~\cite{freeferm}.  Models in this class~\cite{faraggi,chl} have the
ingredients of the MSSM: $N=1$ SUSY,
the SM gauge group as part of the gauge structure, and candidate fields for three
ordinary families and two electroweak Higgs doublets.  Such models typically
have an extended gauge structure which includes an anomalous $U(1)_A$, a number of
non-anomalous $U(1)$'s, and a non-Abelian hidden sector gauge group. The models
also contain a large number of additional matter fields, often with exotic
SM quantum numbers.
The superpotential is calculable in principle to all
orders in the nonrenormalizable terms, and has the feature that string selection
rules can forbid terms allowed by gauge invariance.  
After introducing the required soft supersymmetry breaking parameters, the
phenomenological implications of these models can be investigated.

The standard anomaly cancellation mechanism~\cite{dsw}
leads to the generation of a Fayet-Iliopoulos (FI) term $\xi = g^2_{\rm
str}M_P^2 \Tr Q_A/192\pi^2$ to the $D$-term of $U(1)_A$ at genus-one
\footnote{Here $g_{\rm str}=g/\sqrt{2}$, and $M_P=M_{\rm Planck}/\sqrt{8\pi}$,
with $M_{\rm Planck} \sim 1.2 \times 10^{19}\,GeV$.}. The FI
term induces certain scalar fields to acquire string-scale VEV's along $D$- and
$F$- flat directions, leading to a  restabilized supersymmetric string vacuum.
The vacuum shift reduces the rank of the gauge group, and leads to the generation
of effective mass terms and trilinear couplings from higher-order terms after
replacing the fields in the flat direction by their VEV's.   
The classification of the flat
directions of a general perturbative heterotic string model was
addressed in~\cite{cceel2,cceel3}, and the techniques for computing the
effective superpotential in~\cite{cceelw}.  

\section{Flat Directions}
For the sake of simplicity, consideration is 
restricted to the flat directions involving the (hypercharge preserving)
fields which are singlets under the non-Abelian groups of the model.
The flatness conditions are given by
\bea
\label{anomd}
D_{\rm A} &=& \sum_i Q^{(A)}_i |\varphi_{i}|^2
+ \xi = 0\\
\label{nanomd}
D_{a} &=& \sum_{i} Q^{(a)}_{i}|\varphi_{i}|^2 = 0\\
\label{fflat}
F_{i} &=&\frac{\partial W}{\partial \Phi_{i}} = 0;
\,\, W  =0,
\eea
in which the index $a$ labels the non-anomalous $U(1)$'s.

The method~\cite{cceel2}
utilizes the correspondence between $D$- flat  directions and holomorphic
gauge invariant monomials (HIM's)~\cite{hims}.  The superbasis of all 
independent one-dimensional 
HIM's under the nonanomalous $U(1)$'s is constructed. The elements
with anomalous charge opposite in sign to $\xi$ are also $D_A$ flat for
particular values of the VEV's, set by the FI term to be
$\sim 0.01M_{\rm Planck}$. These elements are the 
building blocks of the complete set of $D$- flat directions of the model\footnote{In more complicated
flat directions, there can be some VEV's that remain undetermined after imposing
the $D_A$-flatness constraint (\ref{anomd}).}. 

The $D$- flat directions are then tested for $F$- flatness.  There are
two types of dangerous terms in the superpotential that can lift a given flat
direction $P$.  First, there are terms which involve only the fields in $P$:
\begin{equation}
\label{wa}
W_A\sim\left(\Pi_{i\in P} \Phi_i\right)^n,
\end{equation}
and there are also the terms linear in an additional field $\Psi$ not in the flat
direction:
\begin{equation}
\label{wb}
W_B\sim\Psi\left(\Pi_{i\in P} \Phi_i\right).
\end{equation}

A flat direction for which gauge invariance allows terms of type-A (\ref{wa}) will
remain $F$- flat only if  string selection rules
conspire to forbid the infinite number of $W_A$ terms, which is difficult to prove
in general.  In contrast, flat direction exist for which gauge invariance only
allows type-B terms (\ref{wb}).
Such type-B flat directions  can be proved to be $F$- flat to all orders in the
nonrenormalizable terms (and to all orders in the string genus
expansion)~\cite{cceel2}, by first constructing the
finite number of type-B superpotential terms (using the requirements of gauge
invariance), then doing a
string calculation to verify if the terms are present (or are forbidden by
world-sheet selection rules).  

Therefore, consideration is restricted to the type-B directions
(although in doing so,  the possible type-A directions which are truly
$F$- flat are missed).
The classification of the type-B flat directions has been carried
out~\cite{cceel2,cceel3} for the free fermionic models of \cite{faraggi,chl}.
For the models considered, in general at least one
additional $U(1)'$ as well as $U(1)_Y$ is left unbroken~\footnote{For a review of $Z'$
physics in string models, see ~\cite{pgl}.}.  

\section{Effective Couplings}
For a given flat direction $P$, effective mass terms can be generated for the
fields $\Psi_i$, $\Psi_j$ via 
\begin{equation}
\label{eq:effmass}
W\sim\Psi_i\Psi_j\left(\Pi_{i\in P}\Phi_i\right)\ .
\end{equation}
The fields with effective mass terms will acquire string-scale masses
and decouple from the theory~\cite{kg,ceew}.   
In addition to the Yukawa couplings of the original superpotential, effective
trilinear interactions for the light fields may also be generated via 
\begin{equation}
\label{eq:efftril}
W\sim\Psi_i\Psi_j\Psi_k\left(\Pi_{i\in P} \Phi_i\right)\ .
\end{equation}
The method for computing these terms is analogous to that of the determination of
the type-B superpotential terms.  First, the complete set of bilinear and
trilinear invariants under the unbroken gauge group is constructed, and gauge
invariance is used to determine the possible couplings of such terms to the
fields in the flat direction. An explicit string calculation then determines which
couplings are present in the superpotential.   

The effective trilinear couplings are typically 
suppressed\footnote{See \cite{units} for a clarification of the determination of
the coupling strengths in a general perturbative heterotic string model.} relative
to the trilinear couplings of the original superpotential (which are $\sim{\cal
O}(g)$). Therefore, there are implications for low energy physics (such as
a possible origin/explanation of the fermion mass hierarchy)
.  

The analysis is currently under investigation~\cite{cceelw} for a number of the
type-B flat directions of a prototype string model. Although there is no
expectation such models will be fully realistic, the analysis sets the stage to
address the generic phenomenological implications of this class of string models.

\section*{Acknowledgments}
I thank G. Cleaver, M. Cveti\v{c}, J. R. Espinosa, P. Langacker, and J. Wang for
enjoyable collaborations on the topics presented.

\end{document}